\documentclass[amsmath,amssymb,floatfix,twocolumn,superscriptaddress]{revtex4}
\usepackage{graphicx}
\usepackage{epsfig}
\usepackage{amsmath}
\usepackage{color}

\begin{document}

\title{Symmetry-enforced heavy-fermion physics in the quadruple-perovskite CaCu$_{3}$Ir$_{4}$O$_{12}$}

\author{Min Liu}
\affiliation{Beijing National Laboratory for Condensed Matter Physics and
Institute of Physics, Chinese Academy of Sciences, Beijing 100190, China}
\affiliation{College of Physical Science and Technology, Sichuan University, Chengdu 610065, China}
\author{Yuanji Xu}
\author{Danqing Hu}
\affiliation{Beijing National Laboratory for Condensed Matter Physics and
Institute of Physics, Chinese Academy of Sciences, Beijing 100190, China}
\affiliation{School of Physical Sciences, University of Chinese Academy of Sciences, Beijing 100049, China}
\author{Zhaoming Fu}
\email[]{fuzm.phy@htu.edu.cn}
\affiliation{Beijing National Laboratory for Condensed Matter Physics and
Institute of Physics, Chinese Academy of Sciences, Beijing 100190, China}
\affiliation{College of Physics and Material Science, Henan Normal University, Xinxiang 453007, China}
\author{Ninghua Tong}
\affiliation{Department of Physics, Renmin University of China, Beijing 100872, China}
\author{Xiangrong Chen}
\email[]{xrchen@scu.edu.cn}
\affiliation{College of Physical Science and Technology, Sichuan University, Chengdu 610065, China}
\author{Jinguang Cheng}
\affiliation{Beijing National Laboratory for Condensed Matter Physics and
Institute of Physics, Chinese Academy of Sciences, Beijing 100190, China}
\affiliation{School of Physical Sciences, University of Chinese Academy of Sciences, Beijing 100049, China}
\author{Wenhui Xie}
\affiliation{Department of Physics, Engineering Research Center for Nanophotonics and
Advanced Instrument, East China Normal University, Shanghai 20062, China}
\author{Yi-feng Yang }
\email[]{yifeng@iphy.ac.cn}
\affiliation{Beijing National Laboratory for Condensed Matter Physics and
Institute of Physics, Chinese Academy of Sciences, Beijing 100190, China}
\affiliation{School of Physical Sciences, University of Chinese Academy of Sciences, Beijing 100049, China}
\affiliation{Collaborative Innovation Center of Quantum Matter, Beijing 100190, China}


\maketitle 

\textbf{Heavy-fermion materials are mostly rare-earth or actinide intermetallics with very few exceptions in $d$-electron systems  \cite{Stewart1984,Hill1970,Coleman2007,Hewson1993,Imada1998}. The physical mechanism for these $d$-electron heavy fermion systems remains unclear \cite{Kondo1997,Arita2007,Shimizu2012,Tomiyasu2014,Ballou1996,Mekata2000,Kobayashi2004,Tanaka2009,Mukherjee2012,Cheng2013,Meyers2014,Wu2016}. Here by studying the quadruple-perovskite CaCu$_3$Ir$_4$O$_{12}$ \cite{Cheng2013}, we propose a symmetry-based mechanism that may enforce heavy-fermion physics in $d$-electron systems. We show that electron hoppings between neighboring Cu 3$d$-orbitals are strictly prohibited by the crystal symmetry, so that Cu 3$d$-electrons can only become delocalized through hybridization with other more itinerant bands, resembling that in typical heavy-fermion rare-earth intermetallics. This provides a useful way to enforce heavy-fermion physics in $d$-electron systems and may help future design of new heavy-fermion materials.}

Heavy-fermion materials are typically rare-earth or actinide intermetallics in which fermionic quasiparticles can have an effective mass hundreds of times higher than the free electron mass \cite{Stewart1984}. One of the primary causes of the heaviness is encoded in the so-called Hill limit, $r_H$, defined as twice the radius of the $f$-orbitals \cite{Hill1970}. When the distance between two neighboring rare-earth or actinide ions is greater than $r_H$, direct overlap between their $f$-orbital wave functions is small, so the low-energy physics is governed by the hybridization between more localized $f$-electron orbitals and other more itinerant orbitals as described in the periodic Anderson model \cite{Coleman2007}. This generates flat hybridization bands near the Fermi energy through the Kondo-like mechanism and the associated quasiparticles are of dominant $f$-characters with a huge effective mass inversely proportional to the narrow bandwidth of the flat bands \cite{Hewson1993}. 

By contrast, this extraordinary property rarely appears in $d$-electron systems. Taking cuprates or iron arsenides as examples, integrating out the high energy degrees of freedom (typically O or As $p$-orbitals) could induce large effective hopping between neighboring $d$-orbitals. As a consequence, these systems are often described by a (multi-orbital) Hubbard model showing no heavy-fermion properties \cite{Imada1998}. However, there do exist few exceptions, including LiV$_2$O$_4$ \cite{Kondo1997,Arita2007,Shimizu2012,Tomiyasu2014}, the Sc-doped YMn$_2$ \cite{Ballou1996,Mekata2000}, the quadruple perovskites ACu$_3$B$_4$O$_{12}$ (A=Ca, La, Na; B=Ru, Ir) \cite{Kobayashi2004,Tanaka2009,Mukherjee2012,Cheng2013,Meyers2014}, and possibly also CsFe$_2$As$_2$ \cite{Wu2016}. These compounds display heavy-fermion properties whose physical origin cannot be easily understood and has stimulated intensive debates ever since their discoveries.

Here we focus on a particular member, CaCu$_3$Ir$_4$O$_{12}$, which exhibits the most pronounced heavy-fermion properties among its family, with a zero temperature specific heat coefficient reported to be 173 mJ/mol K$^2$  \cite{Cheng2013}, higher than that in its sister compound, ACu$_3$Ru$_4$O$_{12}$ \cite{Tanaka2009}. Based on strongly correlated electronic structure calculations, we propose a symmetry-based mechanism for achieving heavy-fermion properties in this family. We show that despite the small nearest-neighbor (NN) and next-nearest-neighbor (NNN) distances between Cu-ions, their effective $d$-electron hoppings (after integrating out the O $p$-orbitals) are strictly prohibited by its special crystal symmetry. The Hill rule is therefore relaxed, suggesting a route that may be useful for future design of new heavy-fermion compounds. 

\begin{figure*}[t]
\begin{center}
\includegraphics[width=0.9\textwidth]{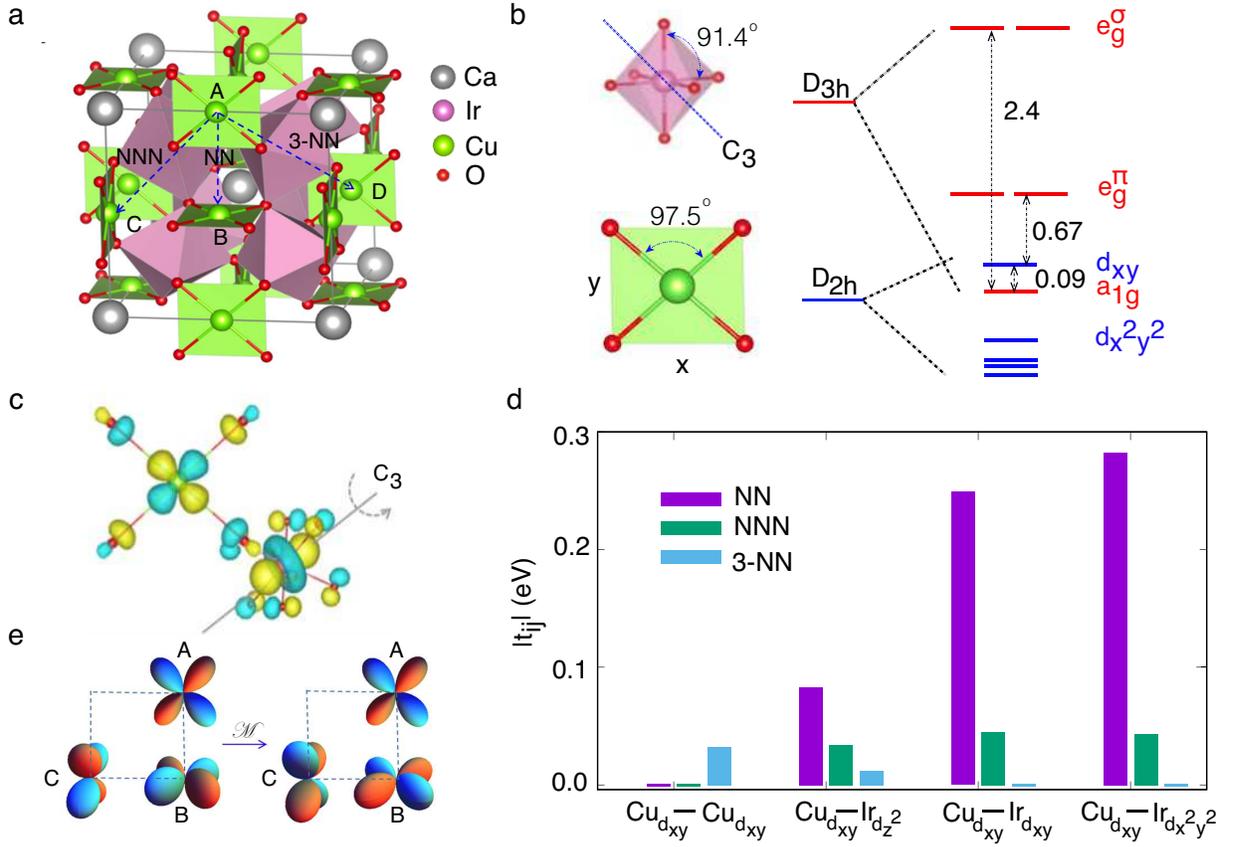}
\caption{\textbf{Crystal symmetry and tight-binding parameters.} \textbf{a,} Schematic crystal structure of CaCu$_{3}$Ir$_{4}$O$_{12}$. The CuO$_{4}$ square planes and the IrO$_{6}$ octahedra are presented by green and pink colours, respectively. The nearest-neighbor (NN), next-nearest neighbor (NNN), and next-next-nearest neighbor (3-NN) Cu-ions are also indicated by dashed lines. \textbf{b,} Schematic plot of the local crystal field levels of Ir 5$d$-orbitals and Cu 3$d$-orbitals in the distorted IrO$_{6}$ octahedra and CuO$_{4}$ rectangle. \textbf{c,} Illustration of the maximally-localized Wannier functions of the Ir 5$a_{1g}$-orbital and the Cu 3$d_{xy}$-orbital. \textbf{d,} The hopping parameters of NN, NNN and 3-NN for Cu 3$d_{xy}$-Cu 3$d_{xy}$, Cu 3$d_{xy}$-Ir $e^{\pi}_{g}$ ($d_{xy}$, $d_{x^{2}-y^{2}}$) and Cu 3$d_{xy}$-Ir $a_{1g}$ ($d_{z^{2}}$)-orbitals in the constructed tight-binding model. The orbitals are labeled according to their local coordinates. The local $z$-axis is perpendicular to Cu-O plane for Cu and parallel to Ca-Ca bond for Ir. \textbf{e,} Illustration of the mirror operation that changes the sign of the NN and NNN Cu 3$d_{xy}$-orbitals.}
\label{fig1} 
\end{center}
\end{figure*}

As shown in Fig.~\ref{fig1}a, CaCu$_3$Ir$_4$O$_{12}$ crystallizes in a body-centered cubic structure with the space group Im$\overline{3}$. Its crystal field scheme can be determined by the density functional theory (DFT) and is illustrated in Fig.~\ref{fig1}b. Each Cu$^{2+}$ ion forms a rectangular plane together with its four neighboring O$^{2-}$ ions. Due to the $D_{2h}$ symmetry, the degeneracy of Cu 3$d$-orbitals is fully lifted. The highest $d_{xy}$-orbital is roughly half filled, while all other 3$d$-orbitals are fully occupied and have no contribution to the low-energy physics. The Ir-ions locate at the center of the IrO$_6$ octahedra which are slightly distorted along the Ca-Ca direction ($\langle 111\rangle$). Its crystal symmetry is therefore lowered from $O_h$ to $D_{3h}$, so that the Ir 5$d$-orbitals are split into two doublets, $e^{\sigma}_g$ and $e^{\pi}_g$, and one singly degenerate orbital, $a_{1g}$. The highest $e^{\sigma}_g$-orbitals are relatively less important. Hence the low-energy physics of CaCu$_3$Ir$_4$O$_{12}$ is governed by the 3$d_{xy}$-orbital of the Cu-ions and the $e^{\pi}_g$ and $a_{1g}$-orbitals of the Ir-ions.

To derive the effective low-energy Hamiltonian, we construct explicitly the maximally-localized Wannier functions (MLWF) of above four orbitals \cite{Blaha2001,Kunes2010}. Figure~\ref{fig1}c depicts the spatial distribution of the Wannier functions of the Cu 3$d_{xy}$-orbital and its neighboring Ir $a_{1g}$-orbital. We see that the $d_{xy}$-orbital lies within the CuO$_{4}$ plane and extends along the Cu-O bonds, while the $a_{1g}$-orbital centers at the Ir-site and extends along the Ca-Ca axis. Their overlap at the intermediate O$^{2-}$-site suggests an O-2$p$ pathway for the hybridization between Cu and Ir $d$-orbitals \cite{Meyers2014}. Considering that the NN Cu-Cu distance is only about 3.7 \AA, much smaller than the Ce-Ce distance in Ce-based heavy-fermion intermetallics, e.g., 4.43 \AA{} in CeAl$_3$ and 4.66 \AA{} in CeRhIn$_5$, one naturally expects a large NN Cu-3$d$ hopping that would prevent CaCu$_3$Ir$_4$O$_{12}$ to exhibit any heavy-fermion behavior. Moreover, the onsite energy of the Cu 3$d_{xy}$-orbital is found to be about $-0.4\,$eV, much closer to the Fermi energy than typical Ce $f$-orbitals. This indicates that the Cu 3$d_{xy}$-electrons are less localized than the $f$-electrons in typical Ce-based heavy-fermion compounds.

Figure~\ref{fig1}d plots the derived hopping parameters for the NN, NNN, and 3-NN Cu-Cu and Cu-Ir orbitals. We focus on the localized Cu 3$d_{xy}$-orbital and its hybridization with the itinerant Ir 5$d$-orbitals. Surprisingly, while the Cu-Ir hybridization parameters decrease rapidly with increasing distance as expected, the NN and NNN Cu-Cu hoppings are both strictly zero, in contrast to the small but finite value of the 3-NN Cu-Cu hopping. We show now that the prohibition of the effective NN and NNN hoppings originates from the special crystal symmetry of CaCu$_3$Ir$_4$O$_{12}$. The hopping parameters are defined as $t_{ij}=\int d^3\bold{r}\mathcal{W}_i(\bold{r})^*\hat{H}\mathcal{W}_j(\bold{r})$, where $\hat{H}$ is the effective Kohn-Sham Hamiltonian and $\mathcal{W}_i$ is the Wannier function of the Cu 3$d_{xy}$-orbital at site $i$. As illustrated in Fig.~\ref{fig1}e, applying a mirror symmetry operation $\mathcal{M}$ with respect to the plane formed by the three Cu-ions, we have $t_{ij}=\int d^3\bold{r}\mathcal{W}^\prime_i(\bold{r})^*\hat{H}\mathcal{W}^\prime_j(\bold{r})=(-1)^{\delta_i-\delta_j}t_{ij}$,
where $\mathcal{W}^\prime_{i}=\mathcal{M}\mathcal{W}_{i}=(-1)^{\delta_i}\mathcal{W}_i$. Upon this operation, the $d_{xy}$-orbital of the Cu-ion (A) lying within the mirror plane remains unchanged, but those of the NN Cu-ion (B) and the NNN Cu-ion (C) are perpendicular to the mirror plane and change sign because of the odd parity of the 3$d_{xy}$ Wannier orbital. Hence $t_{ij}^{NN}=-t_{ij}^{NN}=0$ and $t_{ij}^{NNN}=-t_{ij}^{NNN}=0$, enforced by the crystal and orbital symmetries. On the other hand, the 3-NN CuO$_4$ planes are in parallel direction, so that $t_{ij}^{3-NN}$ is small but finite owing to their large distance (6.47 \AA). Its value is larger than the 3-NN Cu-Ir hopping because the latter has an even larger distance (8.14 \AA). The Cu 3$d_{xy}$-electrons are therefore nearly localized despite the smaller NN and NNN Cu-Cu distances and the Cu-O-Cu pathway. The special crystal structure provides a basis to bypass the Hill rule and yield heavy-fermion properties in CaCu$_3$Ir$_4$O$_{12}$.

\begin{figure}[t]
\begin{center}
\includegraphics[width=0.5\textwidth]{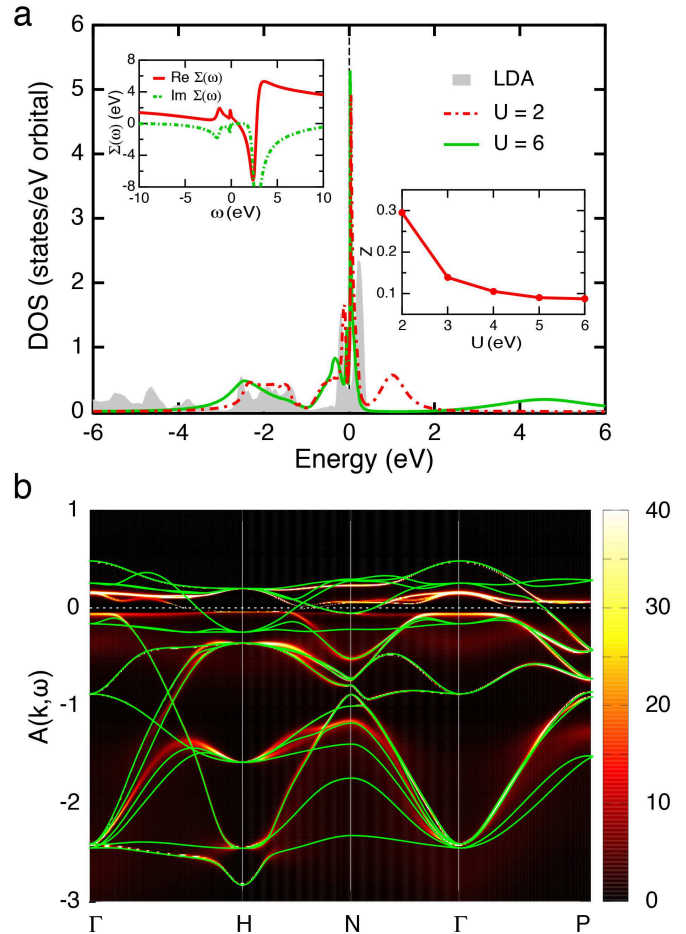}
\caption{\textbf{DMFT density of states and spectral function.} \textbf{a,} DFT+DMFT density of states (DOS) of the Cu 3$d_{xy}$-orbitals using NRG impurity solver for different values of $U$. The shadow denotes the DFT density of states. The insets show the real and imaginary parts of the self-energy $\Sigma(\omega)$ for $U=6\,$eV and the renormalization factor $Z$ as a function of $U$, respectively. \textbf{b,} Momentum-resolved spectral functions $A(k,\omega)$ along the high-symmetry lines in the Brillouin zone obtained by DFT+DMFT(NRG) calculations for $U=6\,$eV compared with the DFT band structures (green solid line).}
\label{fig2}
\end{center}
\end{figure}

To visualize the emergence of heavy electron bands, we further apply the dynamical mean-field theory (DMFT) calculations \cite{Kotliar2006,Held2007}. DMFT maps the derived four-orbital lattice model to a single-impurity model (with a local Coulomb interaction $U$ on the Cu 3$d_{xy}$-orbital) coupled to an effective bath \cite{Georges1996}. Here we solve the impurity model using the numerical renormalization group method \cite{Wilson1975,Bulla2008} and the recently developed continuous-time hybridization expansion quantum Monte Carlo method (CTQMC) \cite{Werner2006,Alps2011}. Figure~\ref{fig2}a compares the DFT density of states of the Cu 3$d_{xy}$-orbital and those obtained by DFT+DMFT(NRG) at zero temperature. The results were further examined using the CTQMC solver (limited to small $U$ and finite temperatures). We see a sharp resonance peak developing slightly above the Fermi energy, indicating a large renormalization of the 3$d_{xy}$-bands due to many-body correlations. The height is further enhanced with increasing $U$. The double peak structure around $\omega=0$ represents the well-known hybridization gap typically seen in heavy-fermion materials. The broad peaks at about 1 eV or 3 eV are from the upper Hubbard band. The self-energy and the resulting renormalization factor $Z=1/(1- \partial \text{Re}\Sigma(\omega)/\partial\omega)|_{\omega=0}$ are plotted in the insets of Fig.~\ref{fig2}a, showing enhanced heaviness, $m^*\sim m_b/Z$, by one order of magnitude at $U=6\,$eV, in addition to the DFT band mass $m_b/m_e\sim 6$, where $m_e$ is the free electron mass. Further enhancement may be obtained from quantum criticality which is beyond the single-impurity DMFT calculations. The development of flat heavy-electron bands can be clearly seen in the momentum-resolved spectral function $A(k, \omega)$ in Fig.~\ref{fig2}b by comparing the DFT and DFT+DMFT results. These bands originate from the Kondo hybridization of the Cu 3$d_{xy}$-orbital with the itinerant Ir-5$d$ orbitals but are of dominant Cu 3$d_{xy}$-character. The DFT results also show hybridization bands because the Cu 3$d_{xy}$-electrons are treated as fully itinerant. In DMFT, these bands are strongly renormalized, shift toward the Fermi energy, and become more flat.

The ACu$_3$B$_4$O$_{12}$ family can adopt a large variety of chemical substitutions, with A=Na, Ca, La, and B=Co, Rh, Ir, Ru, Mn, Fe, Ti, Ge, Sn, etc \cite{Kobayashi2004,Tanaka2009,Mukherjee2012,Cheng2013,Meyers2014,Homes2001,Zeng1999,Mizumaki2011,Shiraki2007}. A systematic survey throughout the whole family reveals further important points. First, the valence of Cu-ions lies always between +2 and +3, so its $d_{xy}$-orbital is the only Cu 3$d$-orbital responsible for the low-energy physics whose NN and NNN hoppings are always prohibited by symmetry. Second, the broad 5$d$-conduction bands of Ir-ions also play an important role. For magnetic B-ions such as Fe and Mn \cite{Zeng1999,Mizumaki2011}, long-range magnetic orders may form to compete with the heavy-fermion behavior. While for nonmagnetic B-ions such as Ti$^{4+}$, Sn$^{4+}$ or Ge$^{4+}$ with either fully filled or nearly depleted outer $d$-shells \cite{Shiraki2007}, heavy-fermion behavior will also be suppressed due to the lack of conduction electrons, causing magnetic orders of the Cu 3$d$-moments. Third, the Cu 3$d_{xy}$-orbital is close to the Fermi energy and its valence can be easily tuned by chemical substitution of either A- or B-ions, driving the system from heavy fermion to mixed valence. These highly adaptive properties make the ACu$_3$B$_4$O$_{12}$ family a fascinating playground for studying the interplay of heavy-fermion physics and other correlated phenomena which are difficult to access in conventional heavy-fermion compounds. It also connects the heavy-fermion intermetallics to the transition metal systems. The symmetry-enforcement mechanism may be applied to general cases and provide a useful pathway for designing new heavy-fermion compounds.

\vspace{20pt}
\noindent
\textbf{Methods}

\noindent The band structures were obtained using the full-potential augmented plane-wave method with generalized gradient approximation (GGA) exchange-correaltion functional as implemented in the WIEN2K code. The Muffin-tin radii ($R_{\text MT}$) are 2.50 a.u. for Ca, 1.90 a.u. for Cu and Ir, and 1.60 a.u. for O. The maximum modulus for the reciprocal vector $K_{\text{max}}$ was chosen such that the smallest $R_{\text MT}*K_{\text{max}}$ is 8.0. All calculations were converged on a grid of 1000 $\bf{k}$-points in the irreducible Brillouin zone. The Wannier orbitals and the tight binding Hamiltonian were constructed based on the maximally localized Wannier function method implemented in the WANNIER90 code. A frozen energy window from -1.2 to 1 eV was chosen for the disentanglement procedure. The resulting 15 bands model include 115-parameters, most of which come from the itinerant Ir 5$d$-bands and are therefore not essential. The DMFT calculations using the CTQMC solver were carried out with the ALPS code. The spectral function was obtained using the maximum entropy analytical continuation method.

\vspace{20pt}
\noindent        
\textbf{Acknowlegements}

\noindent        
This work was supported by the National Natural Science Foundation of China (Grant Nos. 11522435, 11574377, 51572086), the State Key Development Program for Basic Research of China (2015CB921303, 2014CB921500), the Strategic Priority Research Program (B) and Key Research Program of Frontier Sciences of the Chinese Academy of Sciences (Grant Nos. XDB07020200, XDB07020100, QYZDB-SSW-SLH013), the Science Challenge Project (Grant No. JCKY2016212A501) and the NSAF Joint Fund (Grant No. U1430117). Y.Y. was also supported by the Youth Innovation Promotion Association CAS.

\vspace{20pt}
\noindent
\textbf{Author Contributions}

\noindent Y.Y. conceived the idea and supervised the project; M.L., Y.X., D.H., Z.F., and Y.Y. performed the calculations; Y.Y. wrote the paper; all authors contributed to the discussions of the project and the preparation of the paper. M.L. and Y.X. contributed equally to this work.

\vspace{20pt}
\noindent
\textbf{Additional information}

\noindent Correspondence and requests for materials should be addressed to Y.Y., Z.F., or X.C.

\vspace{20pt}
\noindent
\textbf{Competing financial interests}

\noindent The authors declare no competing financial interests.

\end{document}